
\input phyzzx.tex

\def\solid{(---------)}
\def\dashes{($-~-~-~-$)}
\def\dots{($\cdot~\cdot~\cdot~\cdot~\cdot~\cdot\,$)}

\def\dotdash{($\cdot~-~\cdot~-$)}
\def\dotdotdash{($\cdot~\cdot~-~\cdot~\cdot~-$)}


\def\gev{~{\rm GeV}}
\def\tev{~{\rm TeV}}

\def\fbi{~{\rm fb}^{-1}}



\def\prdj#1{{\it Phys. Rev.} {\bf D{#1}}}
\def\npbj#1{{\it Nucl. Phys.} {\bf B{#1}}}
\def\prlj#1{{\it Phys. Rev. Lett.} {\bf {#1}}}
\def\plbj#1{{\it Phys. Lett.} {\bf B{#1}}}

%


\def\wtilde{\widetilde}

\def\etc{{\it etc.}}
\def\ls#1{\ifmath{_{\lower1.5pt\hbox{$\scriptstyle #1$}}}}

%

    \def\fillboxx#1#2{\hbox to #1{\vbox to #2{\vfil}\hfil}
    }

\def\mt{m_t}
\def\mb{m_b}

\def\hl{h^0}
\def\mhl{m_{\hl}}
\def\hh{H^0}
\def\mhh{m_{\hh}}

\def\ifmath#1{\relax\ifmmode #1\else $#1$\fi}
\def\half{\ifmath{{\textstyle{1 \over 2}}}}

\def\quarter{\ifmath{{\textstyle{1 \over 4}}}}
\def\3quarter{{\textstyle{3 \over 4}}}

\def\lplm{l^+l^-}

\def\tanb{\tan\beta}
\def\cotb{\cot\beta}
\def\sinb{\sin\beta}
\def\cosb{\cos\beta}

\def\mw{m_W}
\def\mz{m_Z}

\def\mhp{m_{\hp}}
\def\hp{H^+}
\def\hm{H^-}

\def\rta{\rightarrow}

\def\wp{W^+}
\def\wm{W^-}

\def\hl{h^0}
\def\hh{H^0}
\def\ha{A^0}
\def\mhl{m_{\hl}}
\def\mhh{m_{\hh}}

\def\lam{\lambda}

\def\gam{\gamma}

\def\doeack{\foot{Work supported, in part, by the Department of Energy.}}

\def\gam{\gamma}

\def\VEV#1{\langle #1 \rangle}

\def\tanb{\tan\beta}
\def\sinb{\sin\beta}
\def\cosb{\cos\beta}

\def\wp{W^+}
\def\wm{W^-}
\def\mw{m_W}
\def\mz{m_{Z}}

\def\lam{\lambda}

\def\hp{H^+}
\def\hm{H^-}

\def\epem{e^+e^-}

\def\lplm{l^+ l^-}

\def\rta{\rightarrow}

\def\mt{m_t}
\def\mhp{m_{H^+}}

\def\M2pm{M^2_{P^\pm}}
\def\m2z{\mz^2}

\def\nsd{N_{SD}}
\def\nsdi{\nsd^1}
\def\nsdii{\nsd^2}
\def\nsdiii{\nsd^3}
\def\gaml{\kappa}
\def\sig{\sigma}
\def\fhalfs#1{F^s_{1/2}(#1)}
\def\fhalfp#1{F^p_{1/2}(#1)}
\def\fone#1{F_1(#1)}
\def\fzero#1{F_0(#1)}
\def\dlgamgam{dL_{\gam\gam}}
\def\lame{\lam_e}
\def\zo{\zeta_0}
\def\zi{\zeta_1}
\def\zii{\zeta_2}
\def\ziii{\zeta_3}
\def\zteta{\widetilde\zeta}
\def\zto{\zteta_0}
\def\zti{\zteta_1}
\def\ztii{\zteta_2}
\def\ztiii{\zteta_3}
\def\anti{\overline}
\def\hn{\phi}
\def\hnp{\hn^{\prime}}
\def\hnpp{\hn^{\prime\,\prime}}
\def\mhn{m_\hn}
\def\epsi{{\bf e}}
\def\epsii{{\bf \wtilde e}}
\def\lami{\lam}
\def\lamii{\wtilde\lam}

\def\mpp{M_{++}}
\def\mmm{M_{--}}
\def\mpm{M_{+-}}
\def\mmp{M_{-+}}
\def\absq#1{\left| #1 \right|^2}
\def\im{{\rm Im}\,}
\def\re{{\rm Re}\,}
\def\asym{{\cal A}}
\def\asymi{{\cal A}_1}
\def\asymii{{\cal A}_2}
\def\asymiii{{\cal A}_3}

\Pubnum{UCD-92-18\cr}
\date{June, 1992}
\titlepage
\baselineskip 0pt
\title{Using Back-Scattered Laser Beams to Detect
CP Violation in the Neutral Higgs Sector\doeack}
\author{B. Grz\c{a}dkowski
\foot{On leave of absence from Institute for Theoretical Physics,
University of Warsaw, Warsaw PL-00-681, Poland.\hfil\break
Address after October 1, 1992 : CERN, CH-1211 Geneve 23, Switzerland.}
and J.F. Gunion}
\vskip .15in
\centerline {\it Department of Physics}
  \centerline{\it University of California, Davis, CA 95616}
\vskip .3in
\centerline{\bf Abstract}
\baselineskip 0pt

We demonstrate that the ability to polarize the photons
produced by back-scattering laser beams
at a TeV scale linear $\epem$ collider could
make it possible to determine whether or not a neutral Higgs boson
produced in photon-photon collisions is a CP eigenstate.
The relative utility of different types of polarization is discussed.
Asymmetries that are only non-zero if the Higgs boson is a CP mixture
are defined, and their magnitudes illustrated for
a two-doublet Higgs model with CP-violating neutral sector.

\vskip .4in

An important issue for understanding electroweak symmetry breaking
and physics beyond the Standard Model (SM) is whether or not
there is CP violation in the Higgs sector.  For instance,
both spontaneous and explicit CP violation are certainly possible
in the context of a general two-Higgs-doublet model (2HDM),
whereas CP violation is not possible at tree-level for the specific 2HDM of
the Minimal Supersymmetric Model (MSSM) or its simplest extensions
involving additional singlet scalar Higgs fields.
\Ref\hhg{For a review of Higgs bosons, see
J.F. Gunion, H.E. Haber, G. Kane, S. Dawson, {\sl
The Higgs Hunters Guide}, Addison Wesley (1990).}
The sensitivities of a variety of experimental observables to
CP violation in the neutral sector have been examined
in the literature, ranging from
\REF\weini{S. Weinberg, \prlj{63} (1989) 2333.}
\REF\weinii{S. Weinberg, \prdj{42} (1990) 860.}
\REF\gw{J.F. Gunion and D. Wyler, \plbj{248} (1990) 170.}
\REF\gengng{C.Q. Geng and J.N. Ng, \prdj{42} (1990) 1509.}
\REF\derujulaetal{A. De Rujula, M.B. Gavela, O. Pene and F.J. Vegas,
\plbj{245} (1990) 690.}
\REF\bigi{I. Bigi and N.G. Uraltsev, \npbj{353} (1991) 321.}
\REF\barrzee{S.M. Barr and A. Zee, \prlj{65} (1990) 21.}
\REF\gv{J.F. Gunion and R. Vega, \plbj{251} (1990) 157.}
\REF\texans{R.G. Leigh, S. Paban, and R.-M. Xu, \npbj{352} (1991) 45.}
\REF\changetal{D. Chang, W.-Y. Keung, and T.C. Yuan, \prdj{43} (1991) 14.}
neutron\refmark{\weini-\gw}\ and electron\refmark{\barrzee-\changetal}\
\REF\suzuki{W. Bernreuther and M. Suzuki, {\it Rev. Mod. Phys.}
{\bf 63} (1991) 313.}
electric dipole moments (for a general review and more complete references,
see Ref.~[\suzuki])
to asymmetries in top quark decay and production.
\REF\topas{B. Grzadkowski and J.F. Gunion, preprint UCD-92-7 (1992),
to be published in \plbj{}.}
\REF\schmidtp{C. Schmidt and M. Peskin, preprint SLAC-PUB-5788 (1992).}
\refmark{\topas,\schmidtp}\
However, CP violation in these situations arises via loop graphs
involving the neutral Higgs bosons.
Numerically, current EDM experiments are not sufficiently sensitive
to constrain CP violation in the neutral Higgs sector, although future
results could begin to impose restrictions.
Even if a non-zero EDM is experimentally
observed its interpretation will be uncertain since other types of new
physics could also be involved.
Of course, detection of CP violation in top quark decays and production
must await the large top production rates of the SSC and LHC.

Even after a Higgs boson has been directly observed, it may be difficult
to determine whether or not it is a pure CP eigenstate.
In a CP-conserving 2HDM, there are three neutral Higgs bosons:
two CP-even scalars, $\hl$ and $\hh$ ($\mhl\leq\mhh$) and one
CP-odd scalar, $\ha$. If the 2HDM is CP-violating there will simply
be three mixed states, $\hn_{i=1,3}$. In principle, the presence
of CP violation can be detected through the existence and/or strength
of various couplings. For instance, in the CP-conserving case, at tree-level
the $\ha$ is predicted to have no $WW,ZZ$ couplings while the $\hl$ and $\hh$
together should saturate the $WW,ZZ$ couplings. In the CP-violating case,
all of the $\hn_i$ would have $WW,ZZ$ couplings at tree-level. But, even if
three $\hn_i$ are observed to have $WW,ZZ$ couplings, it would not be
clear whether this was due to CP violation in a 2HDM or to the existence
of more than two doublets. A better possibility is to note
\Ref\pomaral{A. Mendez and A. Pomarol, \plbj{272} (1991) 313.}\
that CP violation at {\it tree-level} would be required if
the couplings $ZZ\hn_1$, $ZZ\hn_2$ and $Z\hn_1\hn_2$ are all non-zero.
To completely avoid contamination from C-violating one-loop
diagrams, three or more neutral Higgs bosons must be detected. Non-zero values
for all three of the couplings $Z\hn_1\hn_2$, $Z\hn_1\hn_3$ and $Z\hn_2\hn_3$
are only possible if CP violation is present.
Finally, we note that the use of correlations between the decay planes of the
decay products of the $WW$ or $ZZ$ vector boson pairs, in order
to analyze the CP properties of the decaying Higgs boson,
\Ref\nelson{C.A. Nelson, \prdj{30} (1984) 1937 (E: {\bf D32} (1985) 1848);
J.R. Dell'Aquila and C.A. Nelson, \prdj{33} (1986) 80,93; \npbj{320}
(1989) 86.}\  will not be useful
in the most probable case where the CP-even component
of a mixed-CP $\hn$ state accounts for essentially all of the $WW,ZZ$
coupling strength.


We wish to contrast the above rather significant difficulties
with the situation that
arises in collisions of polarized photons.  High luminosity
for such collisions is possible using
back-scattered laser beams at a TeV scale linear $\epem$ collider.
\REF\telnovi{H.F. Ginzburg, G.L. Kotkin, V.G. Serbo, and V.I. Telnov,
{\it Nucl. Inst. and Meth.} {\bf 205} (1983) 47.}
\REF\telnovii{H.F. Ginzburg, G.L, Kotkin, S.L. Panfil, V.G. Serbo,
and V.I. Telnov, {\it Nucl. Inst. and Meth.} {\bf 219} (1984) 5.}
\refmark{\telnovi,\telnovii}\
Detection of the SM Higgs boson and of the neutral Higgs bosons of
the MSSM using back-scattered laser beam photons was first studied in
\REF\ghbslaser{J.F. Gunion and H.E. Haber, preprint UCD-90-25
(September, 1990); to appear in the Proceedings of the 1990
DPF Summer Study on High Energy Physics, Snowmass, July 1990.}
Ref.~[\ghbslaser]. More detailed Monte Carlo results have appeared
\REF\bordenetal{D.L. Borden, D.A. Bauer, and D.O. Caldwell,
preprint SLAC-PUB-5715 (January 1992).}
in Ref.~[\bordenetal]. These studies show that expected luminosities
are adequate to make large numbers of Higgs bosons
via such $\gam\gam$ collisions, and that backgrounds to their detection
in $q\anti q$ and $ZZ$ decay channels
are not serious. In $q\anti q$ channels, $\gam\gam\rta q\anti q$ production
can be suppressed in the $m_{Higgs}\gg 2m_q$ limit
by appropriate choices for the helicities of the incoming photons.
The $ZZ$ channel in which one of the $Z$'s decays to $\lplm$ is
virtually background free, and observation of even a handful of events
would constitute an adequate signal. In this paper, we demonstrate that
the ability to control the polarizations of
back-scattered photons provides a powerful means
for exposing the CP properties of any single neutral Higgs
boson that can be produced with reasonable rate. In particular, we find
that there are three polarization asymmetries which are only non-zero
if the Higgs boson is {\it not} a pure CP eigenstate, and which could
well be large enough to be measurable.
We will compute the maximal effects achievable in a general 2HDM
while maintaining a sizeable production rate.

The basic physics behind
our techniques is well-known.\refmark\hhg\  A CP-even scalar couples
to $\gam_1\gam_2$ via $F_{\mu\nu}F^{\mu\nu}$, yielding
(in the center of mass of the two photons)
a coupling strength proportional to $\epsi\cdot\epsii$,
while a CP-odd scalar couples via $F_{\mu\nu}\widetilde F^{\mu\nu}$, implying
a coupling proportional to $(\epsi\times \epsii)_z$.
[Quantities without (with) a tilde belong to $\gam_1$
($\gam_2$).] In the helicity basis,
we employ conventions such that (for $\gam_1$ moving in the $+z$ direction
and $\gam_2$ moving in the $-z$ direction)
$$
e_{\pm}=\mp {1\over \sqrt 2}(0,1,\pm i,0),\qquad
\wtilde e_{\pm}=\mp {1\over \sqrt 2}(0,-1,\pm i,0)\,.\eqn\epsdefs
$$
For these choices we find ($\lami,\lamii=\pm1$):
$$
\epsi\cdot\epsii=-\half(1+\lami\lamii)\,,\qquad
(\epsi\times\epsii)_z= \half i\lami(1+\lami\lamii)\,.\eqn\epsprods
$$
We write the general amplitude for a mixed-CP state $\hn$ to couple to
$\gam\gam$ as $M=\epsi\cdot\epsii\,e+(\epsi\times\epsii)_z\,o$,
where $e$ ($o$) represents the CP-even (-odd) coupling strength.
Then using Eq.~\epsprods\
the helicity amplitude squares and interferences of interest are:
$$
\eqalign{
\absq\mpp+\absq\mmm=&2(\absq e + \absq o)\,,\cr
\absq\mpp-\absq\mmm=&-4\im (eo^*)\,,\cr}
\qquad
\eqalign{
2\re(\mmm^*\mpp)=&2(\absq e - \absq o)\,,\cr
2\im(\mmm^*\mpp)=&-4\re (eo^*)\,.\cr}
\eqn\combos
$$
It is useful to define the three ratios:
$$
\asymi\equiv{\absq\mpp-\absq\mmm \over \absq\mpp+\absq\mmm}\,,
\quad
\asymii\equiv{2\im(\mmm^*\mpp)\over \absq\mpp+\absq\mmm}\,,
\quad
\asymiii\equiv{2\re(\mmm^*\mpp)\over \absq\mpp+\absq\mmm}\,.
\eqn\asymdefs
$$
Note that $\asymi\neq 0$, $\asymii\neq 0$ , and
$|\asymiii|<1$ only if {\it both} the even and odd CP couplings $e$ and $o$
are present. For a CP-even (-odd) eigenstate $\asymi=\asymii=0$
and $\asymiii=+1$ ($-1$).
In the following we demonstrate how to probe these three
ratios, and we will compute their magnitudes in a general CP-violating 2HDM.

The event rate for $\gam\gam$ production of any final state can be
written in the form\refmark\telnovii\
$$dN=\dlgamgam \sum_{i,j=0}^3\VEV{\zeta_i\zteta_j} d\sig_{ij}\,,\eqn\dnform$$
where $\dlgamgam$ is the luminosity for two-photon collisions,
the $\zeta_i$ ($\zteta_j$) are the Stokes polarization parameters
(with $\zo=\zto\equiv 1$) for
$\gam_1$ ($\gam_2$), and $\sig_{ij}$ are the corresponding cross sections.
$\zii$ and $\ztii$ are the mean helicities of the two photons,
while $l=\sqrt{\zi^2+\ziii^2}$ and $\wtilde l=\sqrt{\wtilde
\zi^2+\wtilde\ziii^2}$ are their degrees of linear polarization.
$\dlgamgam$ and $\VEV{\zeta_i\zteta_j}$ are obtained as functions
of the $\gam\gam$ center-of-mass energy, $W$, by averaging over collisions,
including a convolution over the energy spectra for the colliding photons.
\foot{We note that the definition of $\VEV{\ldots}$ employed here
is not the same as that of Ref.~[\telnovii], Eq.~(29), which does
not include a convolution at fixed $W$.  Our definition of $\VEV{\ldots}$
is that implicitly employed in Figs.~13-15 of Ref.~[\bordenetal].}
Expressing the $\sig_{ij}$ in terms of the helicity amplitudes, we obtain
in the case of Higgs boson production:
$$
\eqalign{
dN=\dlgamgam& d\Gamma \quarter(\absq\mpp+\absq\mmm)\biggl\{
(1+\VEV{\zii\ztii})\cr
  +& (\VEV{\zii}+\VEV{\ztii})\asymi +
  (\VEV{\ziii\zti}+\VEV{\zi\ztiii})\asymii
  + (\VEV{\ziii\ztiii}-\VEV{\zi\zti})\asymiii \biggr\}\,,\cr}
\eqn\dnhiggs
$$
where $d\Gamma$ represents an appropriate element of the final state
phase space, including an initial state flux factor.

The behaviors of $\dlgamgam$ and the $\VEV{\zeta_i\zteta_j}$ as functions
of $W$ are crucial to our considerations.
Using standard results for Compton scattering, it is shown in
Ref.~[\telnovii] that both quantities depend sensitively upon
the polarizations of the incoming electron and laser beams.
In order to understand these convolution-weighted quantities, it
will be useful to first discuss the energy spectrum and Stokes
parameters for an individual back-scattered photon, obtained
after integrating only over the azimuthal angles for its emission.
These are determined by four functions.  In particular,
the energy spectrum for $\gam_1$ is directly proportional to
a function $C_{00}$, while its Stokes parameters take the form
$\zeta_i=C_{i0}/C_{00}$, where
$$
\eqalign{
C_{00}=& {1\over 1-y} +1-y-4r(1-r)-2\lame P_c rx(2r-1)(2-y) \cr
C_{20}=&2\lame r x [1+(1-y)(2r-1)^2]-P_c (2r-1)\left[ {1\over 1-y} + 1-y\right]
\cr
C_{10}=& 2r^2 P_t \sin 2\gaml \,, \qquad C_{30}=2r^2P_t\cos2\gaml\,.\cr}
\eqn\cdefs
$$
In Eq.~\cdefs\ $P_c$ ($P_t$) is the degree of circular (transverse)
polarization of the initial laser photon ($P_c^2+P_t^2\leq 1$),
$\gaml$ is the azimuthal angle of the direction of its maximum
linear polarization, and $\lame$ is the mean helicity of the electron
off of which the photon is scattered; note that any sensitivity to
the transverse polarization of the initial
electron has vanished after the azimuthal emission
angle integration. The quantities $r$, $x$ and $y$ are defined by
$$
r={y\over x(1-y)}\,,\quad y={\omega\over E}\,,\quad
x={4E\omega_0\over m_e^2}\simeq 15.3\left({E\over\tev}\right)
\left({\omega_0\over eV}\right)\,,
\eqn\ryxdefs
$$
where $\omega$ ($\omega_0$) is the final (initial) photon energy,
and $E$ is the initial electron energy. The maximum value of
$y$ is $y_{max}=x/(1+x)$, in which limit $r=1$.
All these same quantities as related to the second back-scattered photon
will be denoted with a tilde.

Given that the mass of the Higgs boson is not likely to be near
to the $\epem$ center-of-mass energy, a flat luminosity spectrum
as a function of $W$ is best for Higgs boson searches. This means
that a flat energy spectrum as a function of $y$ is preferred for the
individual photons, and to study the CP properties of the Higgs boson
large values of the $\zeta_i$ ($i=1,2,3$) are required.
To simultaneously achieve both, it turns out that the two extreme choices
of full circular and full transverse polarization for the
initial laser photon are most useful. Consider first $2\lame P_c=\pm1$, \ie\
full circular polarization for the initial laser photon
and maximal average helicity for the incoming electron. For $2\lame P_c=-1$,
$C_{00}$ (and, consequently, the photon spectrum)
is peaked as a function of $y$ just below $y_{max}$, whereas
for $2\lame P_c=+1$ one finds a rather flat (and, hence,
more desirable) spectrum over a broad range
of $y$ falling sharply to 0 as $y\rta y_{max}$.
The behaviors of the $\zeta_i$ follow from Eq.~\cdefs.
For $P_c=\pm1$, $P_t$ must be zero and only $\zii$
(and $\zo\equiv1$) can be non-zero.
The choice of $|2\lame P_c|=1$ allows $\zii$ to be maximal;
one finds $\zii\rta +P_c,-P_c$ for $y\rta 0,y_{max}$,
respectively. In the case of $2\lame P_c=+1$
(preferred for Higgs studies), $\zii\sim +P_c$
over almost the entire $y$ range; only very near to $y=y_{max}$ does
$\zii$ change sign and approach $-P_c$.  This is highly desirable
behavior for isolating $\asymi$.
In contrast, in the case of $2\lame P_c=-1$,
associated with a peaked energy spectrum, $\zii$ slowly switches
sign in the middle of the allowed $y$ range.  Thus, we have a very
fortunate conspiracy in which the $2\lame P_c=+1$ choice which
yields the best photon energy spectrum for study of a Higgs boson, also
yields nearly 100\% circular polarization for the back-scattered photon's
polarization for most $y$ values.

The other extreme choice for the incoming laser beam polarization
is to take $P_c=0$, $P_t=1$.
In this case, $C_{00}$ is independent of $\lame$ and
varies slowly as a function of $y$ over the entire range $y=0$ to
$y=y_{max}$. Eqs.~\dnhiggs\ and \cdefs\
make it apparent that large $l$ will be required in order to
measure $\asymii$ and $\asymiii$.
For $P_t=1$, the linear polarization, $l=\sqrt{\zi^2+\ziii^2}$,
vanishes at $y=0$ ($l\simeq y^2/x^2$) and
is rather small until $y\gsim y_{max}/2$; as $y\rta y_{max}$,
$l$ approaches a maximum of
$l_{max}= 2/ [(1+x)+(1+x)^{-1}]$.  Obviously, to maximize $l$
it would be highly advantageous to have a machine design with as small
a value of $x$ as possible.  It is also useful to note
that if $|2\lame|=1$
then $\zii$ goes from 0 at $y=0$ to a maximum of $|\zii|_{max}=\sqrt{1-l^2}$
at $y=y_{max}$.  For typical values of $x$ (\eg\ of order $2-4$)
$|\zii|_{max}$ can be sufficiently large that it would be useful in suppressing
$q\anti q$ backgrounds.

Of course, to isolate $\asymi$, $\asymii$ and $\asymiii$, it is necessary
to consider the polarizations of both of the back-scattered photons.
When good circular polarization for both laser beams is available,
$\asymi$ would be most easily isolated by
making the wide-spectrum choices of $2\lame P_c=+1$ and
$2\wtilde\lame\wtilde P_c=+1$.  From Eq.~\dnhiggs, we see that
the term proportional to $\asymi$ changes sign if
we reverse the sign of all of the helicities of the incoming electrons
and laser beams --- $\lame$, $\wtilde \lame$, $P_c$ and $\wtilde P_c$ ---
thereby keeping $2\lame P_c$ and $2\wtilde\lame \wtilde P_c$
fixed at $+1$ while reversing the sign of
both $\zii$ and $\ztii$. To determine $\asymii$ and $\asymiii$
we would take $P_t=1$ {\it and} $\wtilde P_t=1$ and note
that the coefficient of $\asymii$ ($\asymiii$) is proportional to
 $\VEV{l\wtilde l}\sin 2(\gaml+\wtilde\gaml)$
($\VEV{l\wtilde l}\cos 2(\gaml+\wtilde\gaml)$).  $\asymii$
could thus be isolated by taking the difference of cross sections
for $\gaml+\wtilde\gaml=+\pi/4$ and $-\pi/4$, while
the difference of cross sections for $\gaml+\wtilde\gaml=0$ and $\pi/2$
would determine $\asymiii$.

A convenient explicit form for the number of Higgs boson events is
obtained by normalizing to the two-photon decay width of the Higgs boson
obtained after summing over final state photon polarizations.
{}From Refs.~[\ghbslaser] and [\bordenetal] the number of Higgs bosons,
$N_\hn$, produced after averaging over colliding photon polarizations is:
$$
\eqalign{
N_{\hn}=&\left.{\dlgamgam\over dW}\right|_{W=\mhn}
{4\pi^2\Gamma(\hn\rta\gam\gam) \over \mhn^2} \cr
\simeq & 1.54\times 10^{4}\left({L_{ee}\over \fbi}\right)
\left({E_{ee}\over TeV}\right)^{-1}\left({\Gamma(\hn\rta\gam\gam)\over
{\rm KeV}}\right) \left({\mhn\over \gev}\right)^{-2}F(\mhn)\,,\cr}
\eqn\nhnform
$$
where $F(W)=(E_{ee}/L_{ee})\dlgamgam/dW$ is a slowly varying function
whose value depends upon details of the machine design, but is ${\cal O}(1)$.
In Eq.~\nhnform, $E_{ee}$ and $L_{ee}$ are the $\epem$ machine energy
and integrated luminosity.  For the case of interest, where
some of the Stokes parameters have non-zero average values,
Eq.~\nhnform\ is modified by the curly bracket
appearing in Eq.~\dnhiggs, with $\zeta_i$
and $\zteta_i$ replaced by $\VEV{\zeta_i}$ and $\VEV{\zteta_i}$, \etc\
All such averages depend upon the $\gam\gam$ invariant mass, $W$.
For instance, for $\asymi=\asymii=0$,
and $P_t=\wtilde P_t=0$, the expressions for $N_{\hn}$ in Eq.~\nhnform\
are multiplied by $(1+\VEV{\zii\ztii}(\mhn))$

Typical behaviors of $F(W)$ and $\VEV{\zii\ztii}(W)$
are amply illustrated in Ref.~[\bordenetal]
for the case of $2\lame P_c\simeq +1$ (\ie\ $P_t=0$) upon which we shall focus.
The most important points to note are the following.
\item{(a)}
A broad spectrum for $F(W)$, advantageous for Higgs studies,
can be achieved using $2\lame P_c$
and $2\wtilde\lame \wtilde P_c$ both as close to $+1$ as possible.
\item{(b)} For $2\lame P_c\sim +1$, $2\wtilde \lame
\wtilde P_c \sim +1$ and $P_c\wtilde P_c\sim +1$,
$\VEV{\zii\ztii}(W)$ is near to $+1$ for $W$ up to 50\%--70\% of $E_{ee}$.
This means that for $\mhn\lsim 70\%E_{ee}$,
Higgs boson production, proportional to $1+\VEV{\zii\ztii}(\mhn)$
(see Eq.~\dnform),
will be enhanced significantly relative to $q\anti q$ backgrounds
which, for small $m_q$, will be suppressed by the factor
$1-\VEV{\zii\ztii}(\mhn)$, as we outline shortly.
\item{(c)} For $P_c\simeq \wtilde P_c\sim \pm1$,
$\VEV{\zii}\sim\VEV{\ztii}\sim \pm 1$ in this same range of $W=\mhn$,
so that $\asymi$ can be easily isolated by simultaneously changing the signs of
$P_c$ and $\wtilde P_c$ for the incoming laser beams (keeping
$2\lame P_c$ and $2\wtilde\lame\wtilde P_c$ fixed near $+1$).

We now turn to estimating the observability of the Higgs boson
and the associated polarization asymmetries.
In our normalization conventions we compute $\Gamma(\hn\rta\gam\gam)$ as:
$$
\Gamma(\hn\rta\gam\gam)={\alpha^2 g^2\over 1024 \pi^3} {\mhn^3\over\mw^2}
\left(\absq e + \absq o\right)\,,
\eqn\gammaform
$$
where
$$e=\sum_i S_{\hn}^i\,,\quad o=\sum_i P_{\hn}^i\,,\eqn\eoforms$$
and $S_{\hn}^i$, $P_{\hn}^i$ represent CP-even, CP-odd triangle contributions
of type $i$.  The only non-zero $P_{\hn}^i$'s derive (at the one-loop
triangle diagram level) from $i=$charged fermion.   Although our computations
include fermion loops from all quarks and leptons, we illustrate by displaying
the expressions for $e$ and $o$ keeping only $b$, $t$, $W$, and $\hp$
triangles.  In this case, we have
$$
\eqalign{
e=&N_c e_t^2 s_{t\anti t}\fhalfs{\tau_t}+N_c e_b^2 s_{b\anti
b}\fhalfs{\tau_b}
+s_{\wp\wm}\fone{\tau_{\wp}}+s_{\hp\hm}\fzero{\tau_{\hp}}\,,\cr
o=&N_c e_t^2 p_{t\anti t}\fhalfp{\tau_t}+N_c e_b^2 p_{b\anti
b}\fhalfp{\tau_b}\,,\cr
}
\eqn\eocomps
$$
where $N_c=3$ for quarks, $e_t=2/3$ and $e_b=-1/3$ are the $t$ and $b$
fractional charges, and $\tau_i\equiv 4m_i^2/\mhn^2$.
The $F_i$'s are those defined in Appendix~C of Ref.~[\hhg].  We remind
the reader that for large $\tau$, $\fone\tau\rta7$, $\fzero\tau\rta -1/3$,
$\fhalfs\tau\rta -4/3$, and $\fhalfp\tau\rta -2$; note in particular
the small size of $\fzero\tau$ in this limit.
The reduced CP-even (scalar, $s$) and
CP-odd (pseudoscalar, $p$) couplings are given by
\foot{To get the correct relative sign between the contributions of
a fermion, $f$, to $e$ and $o$,
it is crucial to note that (in our convention) its reduced $s$ ($p$)
coupling to $\hn$ must be defined as the coefficient of $-gm_f/(2\mw)$ times
$+1$ ($+i\gamma_5$).}
$$
\eqalign{
s_{t\anti t}={u_2\over\sinb}\,,\quad p_{t\anti t}=&-u_3\cotb\,,\quad
s_{b\anti b}= {u_1\over\cosb}\,,\quad p_{b\anti b}=-u_3\tanb\,,\cr
\phantom{s_{t\anti t}={u_2\over\sinb}\,,\quad p_{t\anti t}=}&
s_{\wp\wm}=u_2\sinb+u_1\cosb\,.\cr}
\eqn\spdefs
$$
Reduced couplings for the charged leptons follow those for $b\anti b$,
while in Eq.~\eocomps\ one would have $N_c=1$ and charge $-1$.
In the above, the $u_i$ specify the eigenstate $\hn$ in
the $\Phi_i$ basis of Ref.~[\weinii] (see Ref.~[\topas] for more details).
In a 2HDM, $\sum_i u_i^2=1$, but they are otherwise unconstrained.
Results for the SM Higgs boson correspond to
taking $u_1=\cosb$, $u_2=\sinb$, and $u_3=0$. More generally,
for a CP-even eigenstate
we would have $u_3=0$, while for a CP-odd eigenstate $|u_3|=1$.
We note that the branching ratios for the $\hn$ to decay to
$b\anti b$, $t\anti t$, $\wp\wm$, $ZZ$. \etc\ are determined
using these same reduced couplings by appropriately
weighting the results for CP-even and CP-odd scalars as given in
Appendix B of Ref.~[\hhg]. In Eq.~\spdefs\ we have not given an
expression for $s_{\hp\hm}$.  The most general possibility
is rather complicated, and there is a great deal
of freedom in its magnitude.  However, it is proportional to
$\mw^2/\mhp^2$ and will not be large if $\mhp$ is large, unless the mass of
one of the other neutral Higgs bosons is much larger than $\mhp$.
Because of the very uncertain value of $s_{\hp\hm}$, the reasonable
probability that it will be small, and the fact that it
enters with a relatively small coefficient from $\fzero{\tau_H}$,
we neglect the charged Higgs loop in the numerical estimates to be given later.

Of course, it is also necessary to understand the structure of possible
backgrounds to our Higgs boson signal.  Consider the
case of the $b\anti b$ background with $\mhn\gg 2\mb$.  In this limit,
the amplitudes $\mpp$ and $\mmm$ for $\gam\gam\rta b\anti b$
may be neglected.  Further, parity invariance implies that $\mmp=\mpm$.
We then obtain as the form of the background corresponding to that
of Eq.~\dnhiggs\ for the Higgs boson:
$$
dN=\dlgamgam d\Gamma\half \absq \mpm
\left\{1-\VEV{\zii\ztii}+\VEV{\ziii\ztiii}+\VEV{\zi\zti}\right\}
\,,\eqn\dnbkgnd
$$
Note that if the colliding photons have perfect
circular polarization with $\zii=\ztii=\pm 1$
(in which case $\zi=\ziii=\zti=\ztiii=0$), then
this background from $b \anti b$ production can be essentially eliminated.
\foot{Even if perfect polarization cannot be achieved, $\gam\gam\rta b\anti b$
can be dramatically suppressed by requiring that the
$b$ and $\anti b$ not emerge close to the beam direction.}
Most importantly for determining the CP properties of the $\hn$,
we note that in Eq.~\dnbkgnd\ there are no terms
containing the same $\zeta$/$\zteta$-dependent factors that multiply
$\asymi$, $\asymii$ and $\asymiii$ in Eq.~\dnhiggs.
This remains true even if $\mpp$ and $\mmm$ cannot be neglected
(as, in particular, in the case of the $t\anti t$ background). Thus,
for example, the background cancels when isolating $\asymi$
by simultaneously changing the sign of both $\zii$ and $\ztii$.

We will not attempt to study the backgrounds in detail here.  Instead, we
will use the results of Refs.~[\ghbslaser] and [\bordenetal] to estimate
that a Higgs boson could be seen if its polarization-averaged event rate
is such that there are at least 80 $b\anti b$ decays, or 80 $t\anti t$ decays,
or 20 clean $ZZ$ (with one $Z\rta\lplm$) decays.
Of course, a more detailed Monte Carlo would be required to determine
accurately the minimal signal as a function of $\mhn$ that would be
required in the various channels. For instance, for $b\anti b$ invariant
masses below about $50\gev$, 80 Higgs events in the $b\anti b$
channel would not be adequate.  In addition, the number of background
events depends strongly on the polarization mode employed.
As we have already noted, the $b\anti b$ background can be greatly suppressed
if $2\lame P_c=2\wtilde\lame \wtilde P_c\sim +1$ and $\mhn$ is
below about 70\% of $E_{ee}$ (so that $\VEV{\zii\ztii}(\mhn)\sim 1$).
This is certainly the appropriate configuration for measuring $\asymi$.
On the other hand, the maximal achievable
$\VEV{\zii\ztii}$ for $P_c=\wtilde P_c=0$ (so as to maximize $P_t$ and
$\wtilde P_t$) would not be very near $1$ and full suppression of the
$b\anti b$ background would not be possible
when probing $\asymii$ and $\asymiii$.

For our illustrative results,
we will adopt a machine energy of $E_{ee}=0.5\tev$, an integrated luminosity
of $L_{ee}=20\fbi$, and use the rough value of $F(W)\simeq 1$ in
Eq.~\nhnform\ in computing event rates.
We will search for the values of $|\asymi|$, $|\asymii|$, and $|\asymiii|$
that maximize the observability of CP violation, subject to the minimum
event number requirements stated above.  This search is performed separately
for each of the observables by randomly scanning over all allowed $u_i$ values.
We will present our results as functions of $\mhn$ for various values of
$\mt$ and $\tanb$. We will not consider $\mhn$ smaller than $60\gev$.
For such low $\mhn$, not only would the $b\anti b$
background be large (as noted above), but also it turns out that
existing experimental constraints from $Z$ decays at LEP
require that the $\hn \wp\wm$ coupling must be so suppressed that
a significant number of $\hn$ cannot be made in $\gam\gam$ collisions.
\foot{Recall that the $W$ loop is the most important contributor
to the $\hn\gam\gam$ coupling.}

Regarding the statistical significance, $\nsd$, of
event number differences deriving from the $\asym_{1,2,3}$,
it will be useful to define two standardized
scenarios, the first appropriate for measuring $\asymi$ and the
second for measuring $\asymii$ and $\asymiii$.
Suppose the number of events of a given type for unpolarized photons is $N$.
For measuring $\asymi$ we take $\VEV{\zii}\simeq\VEV{\ztii}=\pm 1$
(this requires $P_c\wtilde P_c\simeq 1$
and $\mhn\lsim 70\% E_{ee}$) and find $2N(1\pm\asymi)$ events.
The statistical significance of the asymmetry fluctuation is then
$\nsdi\equiv\sqrt{2 N}|\asymi|$.  For measuring $\asymii$ and $\asymiii$,
we take $P_t\simeq\wtilde P_t\simeq 1$ and adopt the
``typical'' values of $\VEV {l}\simeq\VEV{\wtilde l}\simeq 0.4$
(not attained until $\mhn\gsim 60\%E_{ee}$ according to our estimates)
and $\VEV{\zii\ztii}=0.5$ (we assume that $2\lame\simeq 2\wtilde\lame\simeq
\pm 1$ so that this latter average is large, in order to suppress $q\anti q$
backgrounds, when $l$ and $\wtilde l$ are large --- see earlier discussion).
The number of events obtained (after averaging over
$2\lame=2\wtilde \lame=\pm 1$) for
$\gaml+\wtilde\gaml=+\pi/4,-\pi/4$ (for $\asymii$) or for
$\gaml+\wtilde\gaml=0,\pi/2$ (for $\asymiii$) is
$N[(1+\VEV{\zii\ztii})\pm \VEV{l\wtilde l}\asym_{2,3}]$.
Using the typical values stated
above we obtain an approximate statistical significance for the event number
fluctuation of $\nsdii\equiv 0.13 \sqrt N |\asymii|$, in the case of
$\asymii$. For $\asymiii$, recall that the signal for CP violation
is $|\asymiii|<1$.  Thus, what matters in this case
is whether the magnitude of fluctuation predicted for
$|\asymiii|=1$ could be distinguished from that associated with some
smaller value of $|\asymiii|$.   The statistical significance
that we shall associate
with this difference will be $\nsdiii\equiv 0.13\sqrt N (1-|\asymiii|)$.

\FIG\asymextremai{}
\FIG\nsigmai{}
\midinsert
\vbox{\phantom{0}\vskip 4.4in
\phantom{0}
\vskip .5in
\hskip -20pt
\special{ insert scr:bslaserbeams_asymextremai.ps}
\vskip -1.4in }
{\rightskip=3pc
 \leftskip=3pc
\Tenpoint
\baselineskip=12pt
\noindent Figure~\asymextremai:
The values for $|\asymi|$ \solid\ and $|\asymii|$ \dashes\
and $(1-|\asymiii|)$ \dots\ which yield the largest
standard scenario statistical significances, $\nsdi$, $\nsdii$,
and $\nsdiii$, respectively (see text), as a function
of $\mhn$.  We have taken $\tanb=2$ and $\mt=150\gev$. Extrema
are obtained for 150,000 random choices of the $u_i$
subject to the requirement that there be at least
80 events in the $b\anti b$ decay channel of the $\hn$,
or 20 events in the $ZZ$ (one $Z\rta \lplm$) channel, or 80
events in the $t\anti t$ channel when the colliding photon polarizations
are averaged over.
}
\endinsert

\midinsert
\vbox{\phantom{0}\vskip 4.4in
\phantom{0}
\vskip .5in
\hskip -20pt
\special{ insert scr:bslaserbeams_nsigmai.ps}
\vskip -1.4in }
{\rightskip=3pc
 \leftskip=3pc
\Tenpoint
\baselineskip=12pt
\noindent Figure~\nsigmai:
The maximum statistical significances $\nsdi$, $\nsdii$ and $\nsdiii$
for observing $|\asymi|$ \solid, $|\asymii|$ \dashes,
and $(1-|\asymiii|)$ \dots, respectively (for the standard scenarios
defined in the text), as a function of $\mhn$.
We have taken $\tanb=2$ and $\mt=150\gev$. Extrema
are obtained for 150,000 random choices of the $u_i$
subject to the requirement that there be at least
80 events in the $b\anti b$ decay channel of the $\hn$,
or 20 events in the $ZZ$ (one $Z\rta \lplm$) channel, or 80
events in the $t\anti t$ channel when the colliding photon polarizations
are averaged over.
}
\endinsert

Results for $\tanb=2$ and $\mt=150\gev$
are presented in Figs.~\asymextremai\ and \nsigmai.
In Fig.~\asymextremai\ we plot the values of $|\asym_{1,2}|$
and $1-|\asymiii|$ which yield the maximal statistical significance
($\nsdi$, $\nsdii$, or $\nsdiii$, as defined above) for
detection of each of these three observables. In all cases,
we impose the minimal event number requirements stated earlier.
In Fig.~\nsigmai\ we plot the corresponding $\nsd$ values
themselves. The best statistical significances (as plotted) are achieved
using the  $b\anti b$ channel
for $\mhn<2\mz$, using the $ZZ$ channel for $2\mz<\mhn<2\mt$,
and using the $t\anti t$ channel for $\mhn>2\mt$.
In all cases, the best $u_i$ choices are such that $u_3$ is large,
thereby guaranteeing a significant contribution to the CP-odd amplitude,
$o$. Also, as discussed in more detail below,
in any particular region of $\mhn$, the $u_i$ that maximize
the $\nsd$'s have certain preferred relative signs.  Otherwise,
the $u_i$ choices which yield the extrema plotted are unnoteworthy. Fine tuning
is not required to obtain $\nsd$ values close to those illustrated.
We have also verified that fine-tuning of the parameters of the 2HDM potential
(see Ref.~[\weinii], Eq.~(55)) is not necessary to attain
these $u_i$ choices. This remains true even if we demand that $\hn$
is the Higgs eigenstate of lowest mass {\it and} that the contribution
of the $\hp$ loop to the $\gam\gam\hn$ coupling is negligible.
Of course, if $\hn$ is not the lightest eigenstate, then decays
such as $\hn\rta\hnp\hnpp$, $\hn\rta\hnp Z$, \etc\ become possible,
and can easily be dominant.  Since the CP asymmetries we consider
are defined by the initial photon polarizations, presumably these
alternative final states would also allow measurement of the associated
production rate differences.

According to Fig.~\nsigmai, at least one of
the asymmetries could be observable throughout almost the entire $\mhn$
range studied. However, we must keep in mind that the $\nsd$ values
presented, based on the polarization averages assumed for the
standard scenarios, are overestimated in certain mass regions. In particular,
in the case of $\asymi$, for $\mhn>2\mt$ $\VEV{\zii}\simeq\VEV{\ztii}=\pm 1$
may not be achievable for $E_{ee}=0.5\tev$, since $\mhn>70\% E_{ee}$ for much
of
this range. In the case of $\asymii$ and $\asymiii$,
the large linear polarizations characterizing our
standard $\asymii,\asymiii$ scenario can only be realized for
$\mhn\gsim 60\% E_{ee}$. However, in practice this does not
appear to be an important restriction, since viable statistical
significances for detection of CP violation via these latter two asymmetries
are mostly attained for $\mhn>2\mt$, in any case.
Thus, determination of the CP properties of the $\hn$
via $\asymii$ and $\asymiii$ will be confined to the $\mhn>2\mt$ region.

Let us now discuss the asymmetry observables corresponding to these
maximal $\nsd$ values, see Fig.~\asymextremai. We focus first on $\asymi$.
$|\asymi|$ is generally small and decreases with increasing
$\mhn$ for $\mhn<2\mw$. This is because the only significant imaginary
contribution to the sum of loop amplitudes comes from the $b$ loop,
and this imaginary part declines in magnitude as $\tau_b\ln(4/\tau_b)$.
In this region, maximal statistical significance is achieved
for values of $u_3\gsim 0.6$, and if $u_1$ and $u_2$ have the same
sign. The latter implies that the $\hn WW$ coupling is not small,
thereby keeping the basic production rate significant
(recall that the $W$ loop generally dominates). Once $\mhn$ is above
the $\wp\wm$ threshold a much larger imaginary part
for the CP-even amplitude, $e$, is possible.  In order to maximize
interference, preferred values of $u_3$ remain large. Further, $u_1$ and
$u_2$ continue to have the same sign; this allows for a large
$W$ loop imaginary part, large basic production cross section,
and large $\hn\rta ZZ$ branching ratio.  Not surprisingly, the best
signal is in the $ZZ$ channel for $2\mz<\mhn<2\mt$.
As $\mhn$ passes beyond $2\mt$, for this case where $\mt$ is significantly
larger than $\mw$ it is easier to achieve a large
$t$-loop imaginary part than $W$-loop imaginary part.  The best
statistical significance is achieved if $u_1$ and $u_2$ have opposite signs,
so as to suppress the $\hn WW$ coupling.
This has a twofold effect: the imaginary part of the $t$ loop is
enhanced relative to the overall amplitude magnitude (enhancing the
level of interference) and $\hn\rta t\anti t$ decays can be dominant,
thereby allowing for the most significant signal to be found in
the $t\anti t$ channel.

\FIG\nsigmaii{}
\FIG\asymextremaii{}

\midinsert
\vbox{\phantom{0}\vskip 4.4in
\phantom{0}
\vskip .5in
\hskip -20pt
\special{ insert scr:bslaserbeams_nsigmaii.ps}
\vskip -1.4in }
{\rightskip=3pc
 \leftskip=3pc
\Tenpoint
\baselineskip=12pt
\noindent Figure~\nsigmaii:
The maximum statistical significances for observing
$|\asymi|$, $\nsdi$ (as defined in the text), as a function of $\mhn$.
Various values for $\mt$ and $\tanb$ are considered:
$\tanb=2,\mt=200\gev$ \solid, $\tanb=2,\mt=100\gev$ \dashes,
$\tanb=10,\mt=200\gev$ \dotdotdash, $\tanb=10,\mt=150\gev$ \dotdash,
and $\tanb=10,\mt=100\gev$ \dots.
Extrema are obtained as described for Fig~\nsigmai.
Curves terminate when the minimal event number requirements can
no longer be met.}
\endinsert

\midinsert
\vbox{\phantom{0}\vskip 4.4in
\phantom{0}
\vskip .5in
\hskip -20pt
\special{ insert scr:bslaserbeams_asymextremaii.ps}
\vskip -1.4in }
{\rightskip=3pc
 \leftskip=3pc
\Tenpoint
\baselineskip=12pt
\noindent Figure~\asymextremaii:
The values for $|\asymi|$ which yield the largest $\nsdi$ values as a function
of $\mhn$.  Various values of $\mt$ and $\mhn$ are illustrated:
$\tanb=2,\mt=200\gev$ \solid, $\tanb=2,\mt=100\gev$ \dashes,
$\tanb=10,\mt=200\gev$ \dotdotdash, $\tanb=10,\mt=150\gev$ \dotdash,
and $\tanb=10,\mt=100\gev$ \dots.
Extrema are obtained as described for Fig.~\nsigmai.
Curves terminate when the minimal event number requirements can
no longer be met.}
\endinsert

For this same case of $\mt=150\gev$ and $\tanb=2$,
$\asymii$ and $\asymiii$ are less promising for CP studies, and
the reasons behind their behaviors are more complex. We discuss only the
$\mhn>2\mt$ region, where reasonably large linear polarization is most
likely to be achieved. Here,
the best $\nsdii$ and $\nsdiii$ values are attained using the
$t\anti t$ decay channel. $u_1$ and $u_2$ are chosen to have appropriately
balanced opposite signs (see the form for $s_{\wp\wm}$ in Eq.~\spdefs) such as
to (simultaneously) severely suppress the $W$ triangle loop contribution to $e$
and the decays $\hn\rta WW,ZZ$. For moderate values of $u_3$, the
CP-even and CP-odd amplitudes can then be made comparable because of the
dominance of the $t$ loop and $|\asymii|$ and $1-|\asymiii|$ take on
values near unity. Values of $\nsdii$ and $\nsdiii$ above 1 are generally
possible. Overall, however, the suppression arising from
the small value of $\VEV{l\wtilde l}\sim 0.16$
that is present {\it ab initio} implies that large statistical
significances for these asymmetries would require much larger luminosity.
For this reason, the remainder of our discussion will focus on $\asymi$.

The achievable statistical significances for $\asymi$ (as well as
the other asymmetries) depend upon $\tanb$ and $\mt$.
In Fig. \nsigmaii\ results for the maximal achievable statistical significance
(subject to minimum event number requirements) for observation of $\asymi$
are given for a variety of other $\tanb,\mt$ choices. The corresponding
values of $|\asymi|$ are given in Fig.~\asymextremaii. We briefly explain the
principle features of these plots.
First, consider keeping $\mt$ fixed at $150\gev$ and
increasing $\tanb$ to 10. Recall that large $\tanb$ suppresses the $t\anti t$
coupling and enhances the $b\anti b$ coupling (see Eq.~\spdefs).
For small $\mhn$, $|\asymi|$ is enhanced compared to small $\tanb$
since the (only) imaginary components of $e$ and $o$ come from the $b$ loop
and are now much larger.  Meanwhile, a large
event rate can be maintained through the $W$ loop contribution to $e$.
However, for $2\mt\gsim \mhn\gsim 2\mw$, even though $\im e$ can be large due
to the $W$ loop, $o$ came mainly from the $t$ loop at moderate $\tanb$,
which loop is now severely suppressed by the large $\tanb$ value.
For $\mhn>2\mt$, $\tau_W$ is getting small and the
$W$ loop contribution to event rate cannot be kept large.  Since the $t$
loop contribution to the $\hn$ production rate and
$B(\hn\rta t\anti t)$ are also both suppressed, it becomes impossible
to maintain minimal event rates in either the $ZZ$ or $t\anti t$ channels.
Thus, detection of $\asymi$ for $\mhn\gsim 2\mw$ becomes difficult
at large $\tanb$.

Keeping $\tanb=2$ and increasing $\mt$ from $150\gev$ to $200\gev$
simply moves the $\mhn=2\mt$ threshold (beyond which the $t$ loop and
$t\anti t$ channel dominate $\asymi$ and $\hn$ decays, respectively)
to a higher value.
The systematics behind the behaviors of $\nsdi$ and $|\asymi|$ are
very much as described above for $\tanb=2$ and $\mt=150\gev$.
Lowering $\mt$ to $100\gev$ at $\tanb=2$ means that for $\mhn\gsim 200\gev$ the
$e$ and $o$ amplitudes can both have large imaginary parts.  Thus,
for $\mhn\sim 200\gev$ both $\nsdi$ and the corresponding
$|\asymi|$ are much larger than for $\mt=150\gev$. However, since
{\it both} $\tau_W$ and $\tau_t$ become small as $\mhn$ increases
further, the $W$- {\it and} $t$-loop contributions to
$e$ and $o$ (and event rates) fall, and
both $\nsdi$ and $|\asymi|$ decline rapidly.
Beyond a certain point ($\mhn\sim 275\gev$) there are simply not enough
$ZZ$ events to satisfy our minimal requirement.

In summary, of the three possible CP-sensitive polarization asymmetries,
$\asymi$ provides the best opportunities
for studying the CP properties of a neutral Higgs boson produced
in $\gam\gam$ collisions of polarized back-scattered laser photons.
A non-zero value for $\asymi$ requires that the $\hn\gam\gam$
coupling have an imaginary part, as well as both CP-even and CP-odd
contributions. For a mixed-CP Higgs boson with $\mhn\lsim 2\mw$, measurement
of $\asymi$ will be easiest if $\tanb$ is large since the $b$
loop, which makes the only large contribution
to the imaginary part for such $\mhn$ values, will be enhanced.
For $\mhn>2\mw$, the required imaginary part is dominated by
the $W$ loop (or $t$-loop if $\mhn$ is also $>2\mt$). Large $\tanb$
makes detection of $\asymi$ in this region more difficult since the
dominant CP-odd contribution derives from the $t$ loop, which will
be suppressed. Nonetheless, it is clear from our analysis that
collisions of polarized back-scattered laser photons will provide a significant
opportunity for determining the CP properties of any neutral
Higgs boson that can be produced with reasonable event rate.
Certainly, a substantial effort should be
made to design a machine with maximal polarization for the
incoming electrons and laser beams and the highest possible luminosity.

\bigskip
\centerline{\bf Acknowledgements}
\medskip

We are grateful to D. Borden, D. Caldwell,
and T. Barklow for discussions regarding the degree of polarization
achievable for the backscattered photons.
One of us (JFG) would like to thank the CERN theory group for
support during the initial stages of this work.  BG would
like to thank U.C. Davis for support during the course of this
research.

\refout
\bye